\documentstyle[aps,preprint,tighten,amsfonts,epsf,epsfig]{revtex}

\newcommand{\Ai}{{\rm Ai}}
\newcommand{\Bi}{{\rm Bi}}
\newcommand{\Ci}{{\rm Ci}^{(+)}}

\begin{document}

\draft

\title{Decay in a uniform field: an exactly solvable model}
\author{R M Cavalcanti\footnote{Email address: rmoritz@if.ufrj.br}}
\address{Instituto de F\'\i sica,
Universidade Federal do Rio de Janeiro, \\
Caixa Postal 68528, 21941-972 Rio de Janeiro, RJ, Brazil}
\author{P Giacconi\footnote{Email address: giacconi@bo.infn.it} and
R Soldati\footnote{Email address: soldati@bo.infn.it}}
\address{Dipartimento di Fisica, Universit\`a di Bologna,\\
Istituto Nazionale di Fisica Nucleare, \\
Sezione di Bologna, 40126 Bologna, Italy}
\date{November 18, 2003}
\maketitle

\begin{abstract}
We investigate the time evolution of the decay (or ionization) probability
of a $D$-dimensional model atom ($D=1,2,3$) in the presence of a uniform
({\it i.e.}, static and homogeneous) background field. The model atom consists
in a non-relativistic point particle in the presence of
a point-like attractive well.
It is shown that the model exhibits infinitely many
resonances leading to possible deviations from the naive exponential decay 
law of the non-decay (or survival) probability of the initial atomic quantum
state. Almost stable states exist
due to the presence of the attractive interaction,
no matter how weak it is. Analytic estimates as well as numerical
evaluation of the decay rates are explicitly given and discussed.
\end{abstract}

\pacs{PACS numbers: 31.70.Hq, 32.60.+i, 03.65.-w}


\section{Introduction}

Exponential decay is a common feature of
many physical processes; in particular, it is the universal
hallmark of unstable systems such as radioactive nuclei. However,
it is known that under very general conditions
quantum mechanics predicts deviations from the exponential
decay within short as well as long time intervals \cite{Chiu}.
As pointed out by Khalfin \cite{Khalfin}, the latter situation
occurs whenever the spectrum of the Hamiltonian $H$ is bounded from below;
in this case, the Paley-Wiener theorem \cite{Wiener} on Fourier
transforms implies that the non-decay or survival amplitude
$A(t;[\psi]):=\langle\psi(0)|\psi(t)\rangle$
necessarily satisfies
\begin{equation}
\int_{-\infty}^{+\infty}\frac{|\ln |A(t;[\psi])| |}{1+t^2}\,dt<\infty\  .
\end{equation}
This condition clearly rules out an exponential decay for $t\to\infty$,
as this would cause the integral above to 
diverge.\footnote{Another measure of decay, used for states initially
confined inside a region ${\cal M}$ ({\it i.e.}, $\psi(x,t=0)$ vanishes
outside ${\cal M}$), is the nonescape probability, defined as 
$P(t):=\int_{\cal M}|\psi(x,t)|^2\,dx$.
Using Schwarz's inequality one can easily show
that $P(t)\ge|A(t)|^2$, so that Khalfin's argument
also rules out the exponential decay of $P(t)$ for $t\to\infty$.}
Explicit calculations in a number of models \cite{Winter}
show that, in fact, there occurs a crossover from
exponential to power law decay when $t\to\infty$.

One may wonder what happens if the Hamiltonian is not bounded 
from below.
At first glance this might appear to be an academic question, since
any realistic Hamiltonian should be bounded from below.
However, such ``unrealistic'' Hamiltonians are 
often found in physics. Some examples are
the decay of a metastable vacuum through
the formation of bubbles of the true vacuum \cite{Voloshin},
the droplet model for first order phase transitions
in statistical physics \cite{Langer}, or the
ionization of an atom by a static electric field \cite{Landau}.
In the latter case Herbst \cite{Herbst} provided a partial
answer to that question. Let $H=-\Delta+V+Fx$ be the
Hamiltonian describing a one-electron atom in a uniform
electric field. If $V(x,y,z)$ is holomorphic in $x$
and $V(x+ia,y,z)$ is bounded and decreases to zero
as $r:=(x^2+y^2+z^2)^{1/2}\to\infty$ for each $a\in{\mathbb R}$,
whereas $\psi$ is an eigenvector of $-\Delta+V$ with negative
eigenvalue, then (for $F\neq 0$)
\begin{equation}
\label{herbst}
\langle\psi|e^{-iHt}|\psi\rangle
=\sum_{\Gamma_j\le\alpha}C_j\,e^{-iE_jt}
+O\left(e^{-(\alpha+\varepsilon)t/2}\right) ,
\end{equation}
where $E_j$ are the resonances of $H$ 
({\it i.e.}, the complex poles of $(E-H)^{-1}$
in the lower half-plane) 
whereas $\Gamma_j:=-2\,{\rm Im}\,E_j$ are their widths and
$\alpha$ and $\varepsilon$ are suitable positive numbers.
Herbst also showed that ${\rm inf}\{\Gamma_j\}>0$,
in such a way that Eq.\ (\ref{herbst}) ensures
the exponential decay of $A(t;[\psi])$ as $t\to\infty$.

The purpose of the present paper is to investigate the
decay law in a very simple (albeit non-trivial) model,
namely, a one-electron atom in which the Coulomb attractive
potential is replaced by a point-like attractive well
--- an idealization of a very short-range
attractive interaction --- and put under the influence of
a static and uniform electric or gravitational field
(for related simple models, see Refs.\ \cite{Yamabe,Dean}).
The model will be studied
in $D=1,2,3$ space dimensions and will be shown to be
exactly solvable in the one- and three-dimensional cases, whereas
in the two-dimensional case it is solvable up to a quadrature.
In spite of its simplicity, this model unravels some remarkable
features that can be actually evaluated in detail and
become worthwhile to be used as a paradigm with respect to
more realistic  situations, without any
substantial change in the basic physical contents.
To this concern, it is known that,
in the absence of the uniform field,
this model exhibits a bound state
--- see for instance Refs.\ \cite{Albeverio,Demkov2,Jackiw}.

It turns out that, once a background uniform field
has been switched on, an infinite number of resonances arise
in this model. In particular, the state vector that corresponds
to the bound state in the absence of the
uniform field is turned into a {\it bona fide}
quasi-stable state for a sufficiently weak external field.
For instance, if the bound state energy is
of the order of 1 eV, the lifetime of the corresponding
quasi-stable state in the presence of the Earth's
gravitational field is much longer than the present
age of the Universe; even in the presence of a
rather strong laboratory static electric field,
its lifetime is long in comparison to the typical time scales
of atomic and condensed matter physics.
On the other hand, very strong external fields are
expected to create non-perturbative deviations from the naive exponential
decay law. This has been observed in previous numerical studies 
of the present model \cite{Geltman2,Arrighini,Elberfeld}
and will be qualitatively explained in this work. 

The paper is organized as follows. In Section \ref{1D} we first
analyze the one-dimensional case,
where a direct one-to-one correspondence takes place
between the strength of the attractive potential well
and the bound state energy. All the main features of
the model are explicitly exhibited and discussed.
In Section \ref{3D} we generalize our investigation to the
two- and three-dimensional cases. 
Here the renormalization procedure is mandatory, in
order to remove the ultraviolet divergences of the
Green's functions. In so doing, the 
bound state energy in the zero-field case achieves a deeper
physical meaning --- it specifies the self-adjoint extension of the
quantum Hamiltonian operator --- whereas the
renormalized coupling parameters become {\em running}
auxiliary quantities. {\it Mutatis mutandis},
all the main physical properties of the one-dimensional
case are essentially recovered. In Section \ref{conclusions} we draw our
conclusions, whilst we defer some technical although
important details to the Appendices.


\section{The one-dimensional model}
\label{1D}

Let us consider the Hamiltonian\footnote{We use atomic units such
that $\hbar=2m=1$.}
\begin{equation}
\label{model}
H=-\frac{d^2}{dx^2}-\lambda\delta(x)-Fx\ ,\qquad\quad\lambda>0\ ,\quad F>0\ ,
\end{equation}
describing a particle interacting with an attractive
$\delta$-potential
and a uniform background field. In the absence of the field
({\it i.e.}, when $F=0$) there is a single bound state with energy
$E_B=-\lambda^2/4$, the corresponding wave function being given by
$\psi_B(x)=(\lambda/2)^{1/2}\exp(-\lambda|x|/2)$.
Once the uniform field is turned on, this bound state
becomes unstable, in the sense that
$A(t;[\psi_B])\to 0$ as $t\to\infty$.
The precise way in which this occurs will be the subject of this Section.


\subsection{Retarded Green's function}
\label{1D1}

The retarded Green's function $G^+(E;x,x')$ is the solution to
the differential equation
\begin{equation}
\label{EqG}
(E-H)\,G^+(E;x,x')=\delta(x-x')\ ,\qquad\quad E\in{\mathbb C}\ ,
\end{equation}
that satisfies the boundary condition
\begin{equation}
\label{BC}
\lim_{|x|\to\infty}\,G^+(E;x,x')=0\quad{\rm for}\quad{\rm Im}(E)>0\ ;
\end{equation}
it is defined for ${\rm Im}(E)\le 0$ by analytic continuation.
The solution to Eq.\ (\ref{EqG}) is known \cite{Demkov2,Elberfeld,Demkov1},
but we shall derive it here for the sake of completeness.

To solve Eq.\ (\ref{EqG}),
let us first consider the case $\lambda=0$;
it can then be rewritten as
\begin{equation}
\label{Green}
\left(\frac{d^2}{d\rho^2}+\rho\right)G_0^+(\rho,\rho')
=F^{-1/3}\,\delta(\rho-\rho')\ ,
\end{equation}
where
\begin{equation}
\label{rho}
\rho:=F^{1/3}\left(x+\frac{E}{F}\right) .
\end{equation}
The solution to Eq.\ (\ref{Green}) that satisfies
the boundary condition (\ref{BC}) is given by
\begin{equation}
G_0^+(\rho,\rho')=a\,\Ai(-\rho)\,\theta(\rho'-\rho)+
b\,\Ci(-\rho)\,\theta(\rho-\rho')\ ,
\end{equation}
where $\Ai(x)$ and $\Ci(x):=\Bi(x)+i\,\Ai(x)$ are Airy
functions \cite{Abramowitz}
and $\theta(x)$ is the Heaviside step function.
The coefficients $a$ and $b$ are fixed by the matching
conditions at $\rho=\rho'$:
\begin{equation}
\label{cont}
G_0^+(\rho'+0,\rho')=G_0^+(\rho'-0,\rho')\ ,
\end{equation}
\begin{equation}
\label{discont}
\partial_{\rho}G_0^+(\rho,\rho')|_{\rho=\rho'+0}
-\partial_{\rho}G_0^+(\rho,\rho')|_{\rho=\rho'-0}=F^{-1/3}\ .
\end{equation}
Solving these equations one finally arrives at
\begin{equation}
G_0^+(\rho,\rho')=-\pi F^{-1/3}\,\Ai(-\rho_{-})\,\Ci(-\rho_{+})\ ,
\label{G0}
\end{equation}
where $2\rho_{\pm}:=\rho+\rho'\pm|\rho-\rho'|$.

In order to obtain $G^+(E;x,x')$ for $\lambda\ne 0$,
we rewrite Eq.\ (\ref{EqG}) as an integral equation:
\begin{eqnarray}
G^+(E;x,x')&=&G_0^+(E;x,x')-\int_{-\infty}^{+\infty}
dy\ G_0^+(E;x,y)\,\lambda\delta(y)\,G^+(E;y,x')
\nonumber \\
&=&G_0^+(E;x,x')-\lambda\,G_0^+(E;x,0)\,G^+(E;0,x')\ .
\label{G}
\end{eqnarray}
Taking $x=0$, solving for $G^+(E;0,x')$,
and reinserting the result into Eq.\ (\ref{G}) yields
the so called Krein's formula \cite{Albeverio}:
\begin{equation}
\label{G1}
G^+(E;x,x')=G_0^+(E;x,x')-\frac{G_0^+(E;x,0)\,G_0^+(E;0,x')}
{g(\lambda,E)}\ ,
\end{equation}
where
\begin{equation}
\label{den}
g(\lambda,E):=\frac{1}{\lambda}+G_0^+(E;0,0)\ .
\end{equation}


\subsection{Resonant-mode expansion of the propagator}
\label{1D2}

{}From $G^+(E;x,x')$ one can obtain the retarded propagator $K^+(t;x,x')$
by a Fourier transformation:
\begin{equation}
\label{Gt1}
K^+(t;x,x')=i\int_{-\infty}^{+\infty}\frac{dE}{2\pi}\
e^{-iEt}\,G^+(E;x,x')\ .
\end{equation}
It turns out that the following bound on
$G^+(E;x,x')$ holds true in the lower 
half-plane\footnote{More precisely, the bound is valid only outside
the sectors $|\arg(E)+2\pi/3|<\delta$ and $-\delta<\arg(E)<0$,
with $\delta>0$ depending on $|E|$. As shown in Sect.\ \ref{1D3}
and Appendix \ref{C}, these regions contain
poles of $G^+(E;x,x')$ with arbitrarily large absolute values, 
where the inequality (\ref{INEQ}) is obviously false. 
One can, however, make $\delta$ arbitrarily small by taking
$|E|$ sufficiently large.}
for $|E|$ sufficiently large (see Appendix \ref{A}): 
\begin{equation}
\label{INEQ}
|G^+(E;x,x')|\lesssim C\,|E|^{-1/2}\,
\exp\left\{|E|^{1/2}(|x|+|x'|)\right\},\qquad|E|\to\infty\ ,
\end{equation}
where $C$ is a suitable constant. 

This bound allows one to close the contour of integration of
(\ref{Gt1}) when $t>0$ with a semi-circle of infinite
radius in the lower half-plane without changing the value
of the integral. Using Cauchy's theorem, one then obtains
the so-called resonant-mode expansion of the propagator
\cite{Moyses,Moshinsky}:
\begin{equation}
\label{Gt2}
K^+(t;x,x')=\sum_{n}e^{-iE_n t}\,\varphi_n(x)\,\varphi_n(x')\ ,
\end{equation}
where the sum runs over the poles\footnote{In writing (\ref{Gt2})
and (\ref{Gamow}) we have made use of the fact that the poles of
$G^+(E;x,x')$ are simple, as it can be explicitly checked
by direct inspection.}
of $G^+(E;x,x')$ located
in the lower half-plane and the functions $\varphi_n(x)$
are given by
\begin{equation}
\label{Gamow}
\varphi_n(x)=\frac{G_0^+(E;x,0)}
{[\,-\partial_E G_0^+(E;0,0)\,]^{1/2}}\Bigg|_{E=E_n} .
\end{equation}
The functions $\varphi_n(x)$ can be recognized as the so-called
{\em Gamow states} \cite{Moyses,Gamow}. 
On the one hand, just like the {\it bona fide} energy eigenfunctions,
they satisfy the
differential equation $H\varphi_n(x)=E_n\,\varphi_n(x)$. On the other hand,
the complex quantities $E_n$ do not correspond to the
eigenvalues of the self-adjoint
Hamiltonian operator and, moreover, the Gamow states
are neither normalizable (not even in the sense of
generalized functions, because they diverge when $x\to\infty$)
nor mutually orthogonal.

Using Eq.~(\ref{Gt2}) and the fact that
\begin{equation}
\psi(x,t)=\int_{-\infty}^{+\infty} dx'\,K^+(t;x,x')\,\psi(x',0)\ ,
\qquad\quad t\ge 0\ ,
\end{equation}
we can recast the non-decay amplitude
$A(t;[\psi]):=\langle\psi|e^{-iHt}|\psi\rangle$ in the form of
a resonant-mode expansion:
\begin{equation}
\label{A(t)}
A(t;[\psi])=\sum_n \widetilde{C}_n\,C_n\,e^{-iE_nt}\ ,
\end{equation}
where
\begin{equation}
C_n:=\int_{-\infty}^{+\infty} dx\,\psi(x,0)\,\varphi_n(x),
\qquad\widetilde{C}_n:=\int_{-\infty}^{+\infty}
dx\,\psi^*(x,0)\,\varphi_n(x) .
\end{equation}
Notice that $\widetilde{C}_n\ne C_n^*$ and
$|\varphi_n(x)|^2\sim\exp(F^{-1/2}\,\Gamma_n\,x^{1/2})$
as $x\to\infty$. It follows therefrom that the wave function $\psi(x,0)$
of the initial state must decrease sufficiently
fast at infinity in order that the coefficients $C_n$, 
$\widetilde{C}_n$ exist.
This condition is fulfilled by $\psi(x,0)=\psi_B(x)$.
This still leaves open the question of whether the series
(\ref{A(t)}) converges. Here we shall {\em assume} that it does, at least in
the $l^2$-topology.


\subsection{Poles of the Green's function}
\label{1D3}

\begin{figure}[htp]
\centerline{\hbox{\epsfig{figure=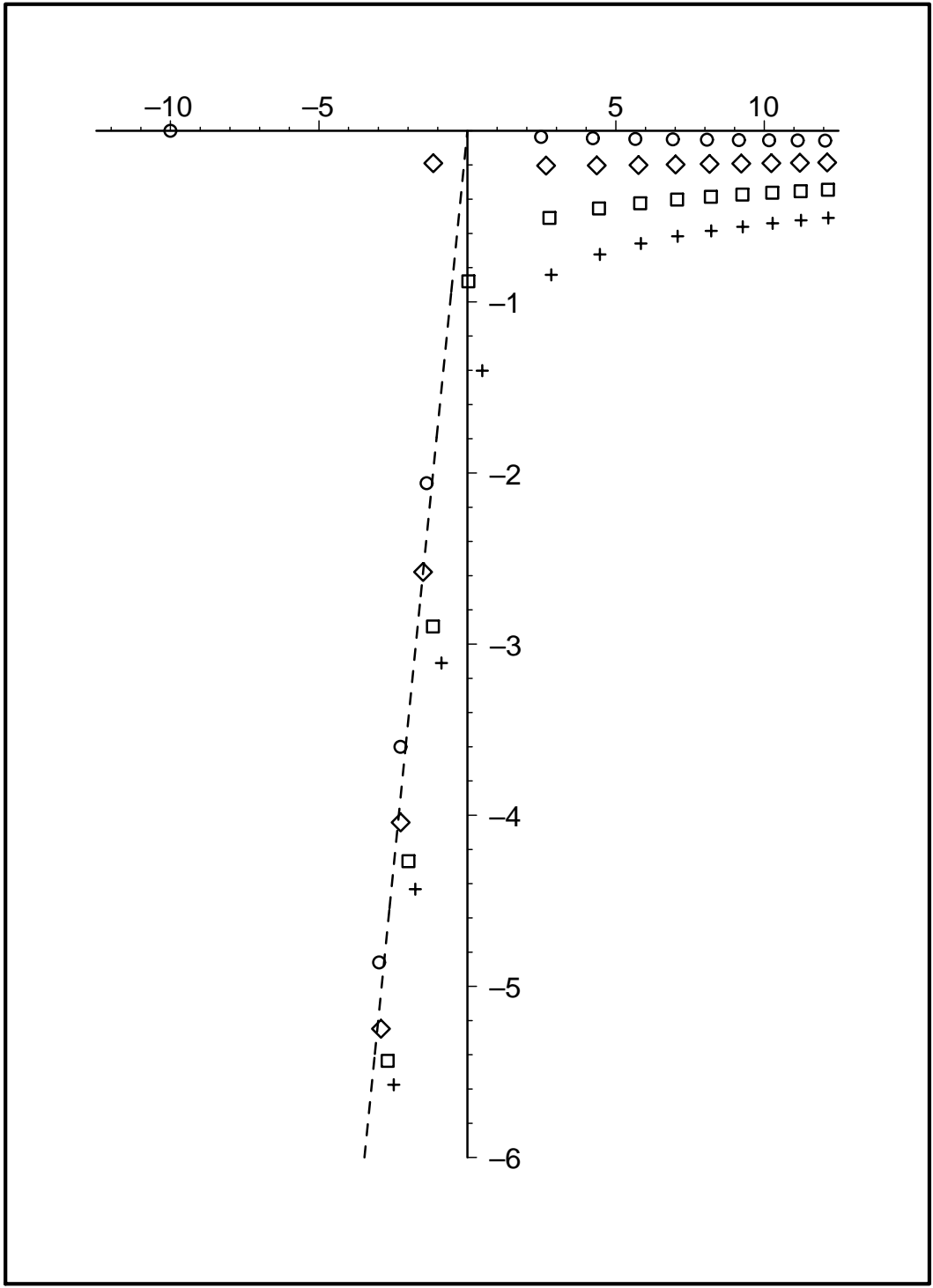,width=14cm}}}
\vspace{0.5cm}
\caption{Poles of the Green's function in the complex $\varepsilon$-plane 
($\varepsilon_{-3}$ to $\varepsilon_9$, clockwise)
in the one-dimensional case: $\varepsilon_B=-10$ ($\circ$),
$\varepsilon_B=-1$ ($\Diamond$), $\varepsilon_B=-0.1$ ($\Box$), and 
$\varepsilon_B=-0.01$ ($+$). The dashed line corresponds to the half-line 
$\arg(\varepsilon)=-2\pi/3$. (Angles appear distorded in this plot 
because the real and imaginary axes have different scales.)}
\label{fig1}
\end{figure}

The unperturbed Green's function $G_0^+(E;x,x')$ is an
holomorphic function of $E$,
so that the poles of $G^+(E;x,x')$ are
all given by the zeros of $g(\lambda,E)$.
Inserting the explicit form of $G_0^+(E;0,0)$ into Eq.\ (\ref{den})
and noting that when $F=0$ there is a bound
state with energy $E_B=-\lambda^2/4$, we arrive at
the following equation:
\begin{equation}
\label{EQ}
\Ai(-\varepsilon)\,\Ci(-\varepsilon)
=\frac{1}{2\pi}\,(-\varepsilon_B)^{-1/2}\ ,\qquad\quad
\varepsilon_B:= E_BF^{-2/3}\ .
\end{equation}
For a given value of $\varepsilon_B$,
Eq.\ (\ref{EQ}) has an infinite number
of solutions, all located in the lower half-plane. Some of them
are shown in Fig.\ \ref{fig1}. They can be numbered according
to their values in the limit $\varepsilon_B\to-\infty$,
which corresponds to a very weak
field ($F\to 0$) or a very strong attractive interaction ($\lambda\to\infty$).
One of the poles approaches the negative real axis
and behaves asymptotically as (see Appendix B)
\begin{equation}
\label{e01d}
\varepsilon_0\sim\varepsilon_B\left\{1+i\exp\left[-\frac{4}{3}\,
(-\varepsilon_B)^{3/2}\right]\right\}\ ,\quad\qquad
\varepsilon_B\to-\infty\ .
\end{equation}
Its real part corresponds to the energy of the (unique) bound state of
the atom in the absence of the uniform field. Its imaginary part
is half the decay rate of the atom via tunneling through the
potential barrier created by the external field.

The other poles approach the zeros\footnote{Note that the r.h.s.\ of
Eq.\ (\ref{EQ}) vanishes in the limit $\varepsilon_B\to-\infty$.}
of $\Ai(-\varepsilon)$,
which are real and located on the positive real axis,
\begin{equation}
\label{elim1}
\lim_{\varepsilon_B\to-\infty}\varepsilon_n=-a_n\ ,\quad\qquad
n\in{\mathbb N}\ ,
\end{equation}
and of $\Ci(-\varepsilon)$,
\begin{equation}
\label{elim2}
\lim_{\varepsilon_B\to-\infty}\varepsilon_{-n}=-a_n\,e^{-2i\pi/3}\ ,
\quad\qquad
n\in{\mathbb N}\ .
\end{equation}
In Eq.\ (\ref{elim1}), $a_n$ denotes the $n$-th zero of $\Ai(z)$;
Eq.\ (\ref{elim2}) follows from the identity
$\Ci(z)=2e^{i\pi/6}\Ai(ze^{2i\pi/3})$ \cite{Abramowitz}.

If the external field is very weak, but nonvanishing --- {\it i.e.},
$|\varepsilon_B|\gg 1$ --- then
the poles $\varepsilon_n$ with $n>0$ exhibit a
small negative imaginary part
(see Fig.\ \ref{fig1}). However, while ${\rm Im}(\varepsilon_0)$ approaches
zero exponentially fast as $\varepsilon_B\to-\infty$, one has
${\rm Im}(\varepsilon_n)\sim(-\varepsilon_B)^{-1}$ in the same limit
(provided $n$ is not very large, see Appendix \ref{B/C}). This means
that the transient effects associated to the poles $\varepsilon_n$ with $n>0$
--- and {\it a fortiori} those ones associated to
$\varepsilon_n$ with $n<0$ --- disappear much faster
than the corresponding effects associated to the
resonance $\varepsilon_0$.

Looking at Fig.\ \ref{fig1}, one can notice that the imaginary parts of
the first few poles $\varepsilon_n$ with $n>0$ have the same order of
magnitude. This explains the short time oscillatory behaviour
of $|A(t;[\psi])|^2$ observed in numerical studies
\cite{Geltman2,Arrighini,Elberfeld}
of the model (\ref{model}) in the weak field regime:
it is a consequence
of the interference among the resonances associated with those poles.
As a matter of fact,
these resonances have a simple physical interpretation: when the external
field $F$ is turned on, it may excite the particle to a state of positive
energy. Once excited, the particle is pushed to the positive
$x$-direction by the field --- recall that we are assuming $F>0$ ---
but it is scattered by the potential $V(x)=-\lambda\delta(x)$.
Because the potential is strongly attractive as $\lambda$ is very
large, the
transmission probability is small, so that the particle can bounce back and
forth many times in the region to the left of the origin before it finally
``jumps over'' the potential well.

Let us now examine the strong field regime $|\varepsilon_B|\ll 1$.
In this case, as shown in Fig.\ \ref{fig1}, the decay rates $\{\Gamma_j|\,
j\in{\mathbb Z}\}$ form a monotonic decreasing sequence,
with $\lim_{j\to\infty}\Gamma_j=0$, as shown in Appendix \ref{C}. Thus,
in contrast with the class of potentials considered by Herbst \cite{Herbst},
there does not exist a slower decaying resonance, which would eventually
dominate the decay process. As a consequence, the decay is not asymptotically
exponential: $\forall\alpha>0$,
$\lim_{t\to\infty}e^{\alpha t}\,|A(t;[\psi])|^2=\infty$.
By the way, strictly speaking this result actually holds true
even in the weak field regime
$|\varepsilon_B|\gg 1$, because $\lim_{n\to\infty}\Gamma_n=0$ regardless
the value of $\varepsilon_B$ (see Appendix \ref{C}). In the weak field case,
however, one should have to wait an extremely long time until a deviation
from the exponential decay $|A(t;[\psi])|^2\sim\exp(-\Gamma_0t)$ became
appreciable. Besides, $|A(t;[\psi])|^2$ would be so small by then that
such a deviation would be practically unobservable.

The crossover from weak to strong field regime occurs at
$\varepsilon_B\sim -1$.
At this value, ${\rm Im}(\varepsilon_0)\approx{\rm Im}(\varepsilon_1)$
(see Fig.\ \ref{fig1}) --- an indication that the two mechanisms of decay
discussed above become equally important. 


\section{The two- and three-dimensional cases}
\label{3D}

\subsection{Retarded Green's function}
\label{3D1}

We can use the same strategy employed in Sec.\ \ref{1D1}
to solve the $D$-dimensional version of Eq.\ (\ref{EqG}), which reads
\begin{equation}
\label{EqG2}
\left[\,E+\nabla^2+\lambda\delta^{(D)}({\bf x})+Fx\,\right]\,
G^+(E;{\bf x},{\bf x}')=\delta^{(D)}({\bf x}-{\bf x}')\ ,
\end{equation}
where ${\bf x}=(x_1,\ldots,x_D):=(x,{\bf r})$ and $E\in{\mathbb C}$.
Thus we can formally write $G^+(E;{\bf x},{\bf x}')$ as in
Eq.\ (\ref{G1}), in which $G_0^+(E;{\bf x},{\bf x}')$ denotes the
solution to Eq.\ (\ref{EqG2}) in the case $\lambda=0$. The 
latter can be written as
\begin{equation}
\label{G03}
G_0^+(E;{\bf x},{\bf x}')=\int\frac{d^{D-1}k}{(2\pi)^{D-1}}\,
e^{i{\bf k}\cdot({\bf r}-{\bf r}')}\,{\cal G}_0^+(E,{\bf k};x,x')\ ,
\end{equation}
where ${\cal G}_0^+(E,{\bf k};x,x')$ satisfies
\begin{equation}
\left(E-{\bf k}^2+\frac{\partial^2}{\partial x^2}+Fx\right)
{\cal G}_0^+(E,{\bf k};x,x')=\delta(x-x')\ .
\end{equation}
This has precisely the form of Eq.\ (\ref{EqG}) with $\lambda=0$
and $E\to E-{\bf k}^2$, the solution to which is given by Eq.\ (\ref{G0}).
Inserting it into Eq.\ (\ref{G03}) we finally obtain
\begin{equation}
\label{G0D}
G_0^+(E;{\bf x},{\bf x}')=-\pi F^{-1/3}
\int\frac{d^{D-1}k}{(2\pi)^{D-1}}\,
e^{i{\bf k}\cdot({\bf r}-{\bf r}')}\,\Ai(-\rho_{-})\,\Ci(-\rho_{+})\ ,
\end{equation}
where now $\rho:= F^{1/3}\,[x+(E-{\bf k}^2)/F]$ and
$2\rho_{\pm}:=\rho+\rho'\pm|\rho-\rho'|$.


\subsection{Renormalization}
\label{3D2}

In contrast with the one-dimensional case, 
the Green's function is
ill-defined at coincident points for $D\ge 2$.
Indeed, after setting ${\bf x}={\bf x}'=0$ in Eq.\ (\ref{G0D})
and performing the angular integration, we obtain
\begin{equation}
\label{Pole1}
G_0^+(E;0,0)=-C_D\,F^{-1/3}
\int_0^{\infty}\Ai(q)\,\Ci(q)\,k^{D-2}\,dk\ ,
\end{equation}
where
\begin{equation}
\label{defCDq}
C_D:=\frac{2^{2-D}\pi^{(3-D)/2}}{\Gamma[(D-1)/2]}\ ,
\qquad\quad q:= F^{-2/3}({\bf k}^2-E)\ .
\end{equation}
Since 
\begin{equation}
\Ai(q)\,\Ci(q)\sim\frac{q^{-1/2}}{2\pi}
\sim\frac{F^{1/3}}{2\pi k}\quad\quad{\rm for}\quad\quad k\to\infty\ ,
\end{equation} 
the integral in Eq.\ (\ref{Pole1}) turns out to be ultraviolet divergent
in $D\ge 2$. In the two- and three-dimensional cases
the divergence can be absorbed
through a redefinition of the coupling parameter $\lambda$.
To do this we follow the same procedure employed in \cite{CGPS}.
Let us introduce a cutoff $\Lambda$ in the upper limit of
integration in Eq.\ (\ref{Pole1}) and add to the resulting
expression the following integral: 
\begin{equation}
I_D(\Lambda,\mu):=\frac{C_D}{2\pi}\int_0^{\Lambda}
\frac{k^{D-2}\,dk}{\sqrt{k^2+\mu^2}}\ ,
\end{equation}
which contains the arbitrary momentum scale $\mu>0$.
At the same time, we subtract $I_D(\Lambda,\mu)$ from
$\lambda^{-1}$ and define the renormalized coupling
parameter $\lambda_R$ as
\begin{equation}
\left[\lambda_R(\mu)\right]^{-1}:=\lim_{\Lambda\to\infty}\left[\,
\lambda^{-1}-I_D(\Lambda,\mu)\,\right] ,
\end{equation}
where it is understood that $\lambda$ depends on $\Lambda$
in such a way that the limit exists. In this way, the denominator
of the  Krein's formula (\ref{G1}) is replaced by an expression
that is finite when the cutoff is removed:
\begin{eqnarray}
\lim_{\Lambda\to\infty}\,g_D(\lambda,E)&=&
\frac{1}{\lambda_R}-\frac{C_D}{F^{1/3}}
\int_0^{\infty}\left[\,\Ai(q)\,\Ci(q)-\frac{F^{1/3}}
{2\pi\sqrt{k^2+\mu^2}}\,\right]k^{D-2}\,dk
\nonumber \\
&:=&g_D(\lambda_R,\mu,E)\ .
\label{renorm}
\end{eqnarray}

In the next two subsections we shall analyze this expression 
separately in $D=2$ and $D=3$ dimensions. It turns out that the latter
is simpler than the former, so we discuss it first.


\subsection{Three-dimensional case}

In $D=3$ the integral in Eq.\ (\ref{renorm}) can be computed
in closed form\footnote{
$\int y_1 y_2\,dx=x y_1 y_2 - y_1' y_2'$ for any two solutions
of Airy's equation $y''-xy = 0$.} yielding ($\varepsilon:=EF^{-2/3}$)
\begin{equation}
g_3(\lambda_R,\mu,E)=\frac{1}{\lambda_R}-\frac{\mu}{4\pi}
-\frac{1}{4}\,F^{1/3}\left[
\varepsilon\,\Ai(-\varepsilon)\,
\Ci(-\varepsilon)+\Ai'(-\varepsilon)\,{\Ci}'(-\varepsilon)\right] .
\end{equation}
Using the asymptotic expressions of the Airy functions for
large argument \cite{Abramowitz}, one can easily show
that in the limit $F\to 0$ the expression above is reduced to
\begin{equation}
\left.g_3(\lambda_R,\mu,E)\right|_{F=0}=
\frac{1}{\lambda_R}-\frac{\mu}{4\pi}+\frac{\sqrt{-E}}{4\pi}\ .
\end{equation}
Thus, provided $\lambda_R>4\pi/\mu$, the quantity $g_3(\lambda_R,\mu,E)$ has
a real zero given by
\begin{equation}
E_B=-\left[\mu-\frac{4\pi}{\lambda_R(\mu)}\right]^2 ,
\end{equation}
which can be identified as the energy of the unique bound state of the system.
It is worthwhile to remark that the bound state energy
is a physical quantity and turns out to be independent
of the arbitrary scale $\mu$. From this physical requirement
one can readily obtain the flow equation for the renormalized
{\em running} coupling parameter:
\begin{equation}
\lambda_R(\mu)=\frac{\lambda_R(\mu_0)}{1+(\mu-\mu_0)
[\lambda_R(\mu_0)/4\pi]}\ ,
\end{equation}
which exhibits asymptotic freedom, {\it i.e.}, $\lambda_R(\mu)\to 0$
as $\mu\to\infty$.

After setting $\varepsilon_B:=F^{-2/3}E_B$
we can rewrite the resonance equation $g_3(\lambda_R,\mu,E)=0$
in the form
\begin{equation}
\label{EQ3}
\frac{1}{\pi}\,(-\varepsilon_B)^{1/2}+\varepsilon\,\Ai(-\varepsilon)\,
\Ci(-\varepsilon)+\Ai'(-\varepsilon)\,{\Ci}'(-\varepsilon)=0\ ,
\end{equation}
that generalizes Eq.\ (\ref{EQ}) to the three-dimensional case.
As in the one-dimensional case, Eq.\ (\ref{EQ3}) has a
solution $\varepsilon_0$ that
tends asymptotically to $\varepsilon_B$ in the weak-field regime
(see Appendix \ref{B}):
\begin{equation}
\label{e03D}
\varepsilon_0\sim
\varepsilon_B\left\{1+\frac{i}{4}(-\varepsilon_B)^{-3/2}
\exp\left[-\frac{4}{3}\,(-\varepsilon_B)^{3/2}\right]\right\}\ ,
\qquad\quad\varepsilon_B\to-\infty\ .
\end{equation}
It has the same physical interpretation of its one-dimensional
counterpart --- see Eq.\ (\ref{e01d}).

\begin{figure}[htp]
\centerline{\hbox{\epsfig{figure=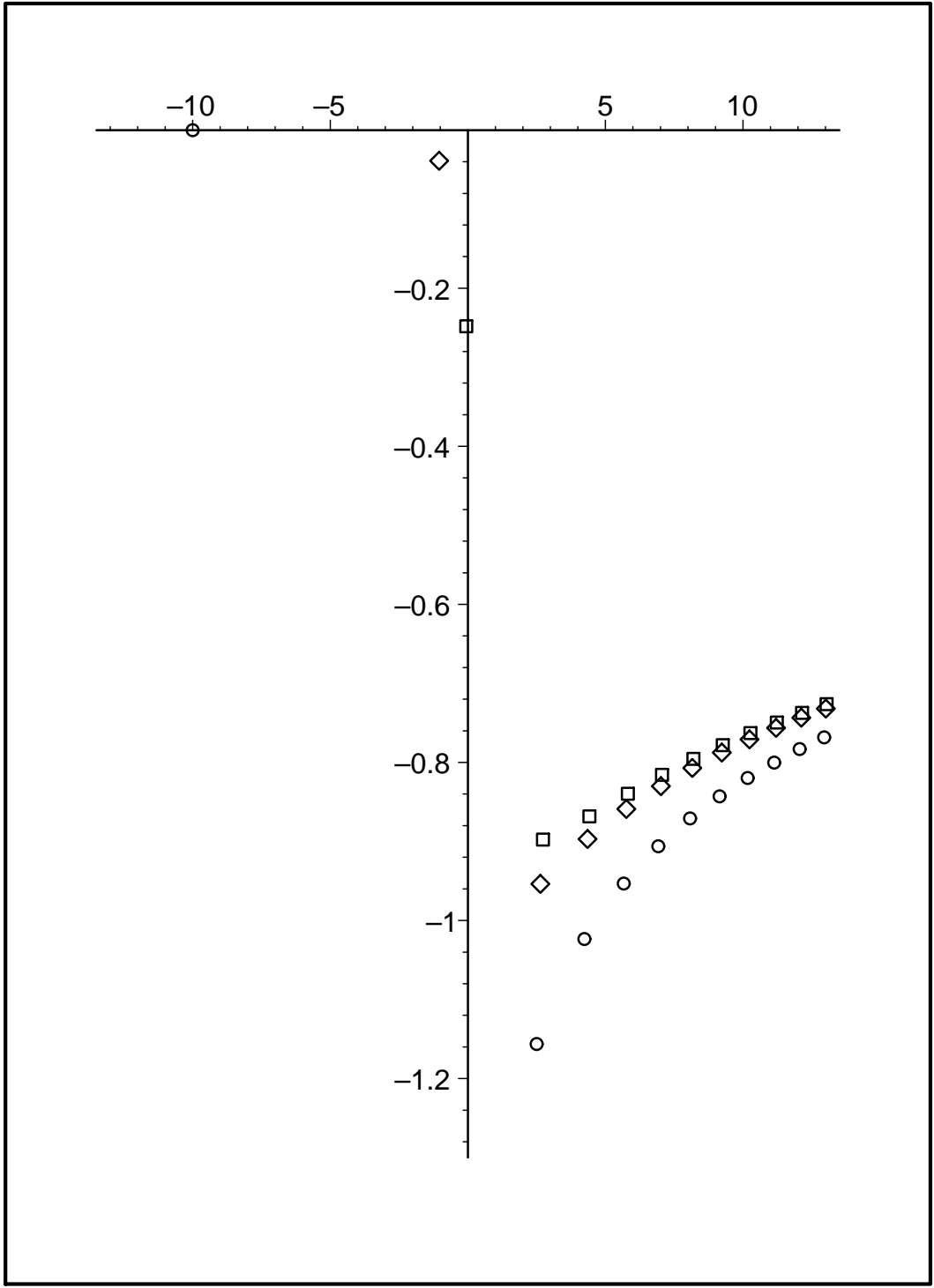,width=14cm}}}
\vspace{0.5cm}
\caption{Poles of the Green's function in the complex $\varepsilon$-plane
($\varepsilon_0$ to $\varepsilon_{10}$, from left to right)
in the three-dimensional case: $\varepsilon_B=-10$ ($\circ$),
$\varepsilon_B=-1$ ($\diamond$), and $\varepsilon_B=-0.1$ ($\Box$).}
\label{fig2}
\end{figure}

In addition to $\varepsilon_0$, Eq.\ (\ref{EQ3}) has an infinite number
of solutions. Some of them are shown in Fig.\ \ref{fig2} for three different
values of $\varepsilon_B$. Their distribution in the complex
$\varepsilon$-plane bears some resemblance with the one-dimensional
case (see Fig.\ \ref{fig1}); in particular, they approach asymptotically
the half-lines $\arg(\varepsilon)=-2\pi/3$ and $\arg(\varepsilon)=0$
(see Appendix \ref{C}). There are, however, two important differences:\\
(i) for fixed $n>0$ we have that\footnote{Note that this fact is not in conflict
with Eq.\ (\ref{asymp3d}), which is valid under the
condition that $|\varepsilon_n|\gg|\varepsilon_B|$.}
$\lim_{\varepsilon_B\to-\infty}{\rm Im}(\varepsilon_n)\neq 0$,
as shown in Fig.\ \ref{fig2}, which clearly exhibits that
the larger $|\varepsilon_B|$ the farther is $\varepsilon_n$ from
the real axis;\\
(ii) there is no clear distinction between the weak and strong field
regimes --- $\Gamma_0$ is always smaller than $\Gamma_1$,
even for $\varepsilon_B\to 0$.

The first difference has a simple geometric interpretation:
in three dimensions a particle can avoid a localized obstacle
by going around it. Hence, the field $F$ can easily detach a particle
with positive energy from a localized potential.


\subsection{Two-dimensional case}
\label{D=2}

Let us now finally discuss the two-dimensional case. To this concern,
it is important to realize that the integral in Eq.\ (\ref{renorm}) 
is no longer expressible in closed form when $D=2$. 
Here we shall content ourselves with deriving an asymptotic
expression for $\varepsilon_0$ in the limit $F\to 0$.
To this aim let us assume that $|\varepsilon_0|$ is large and close to
the negative real half-axis. In this case --- see Eq.\ (\ref{defCDq}) ---
$|q|$ is large and $|\arg(q)|\approx\pi$ for all $k\in[0,\infty)$,
hence the following approximation is uniformly valid in the range of
integration in Eq.\ (\ref{renorm}) \cite{Abramowitz}: 
\begin{equation}
\Ai(q)\,\Ci(q)\sim\frac{F^{1/3}}{2\pi\sqrt{k^2-E}}\left\{1+\frac{i}{2}\,
\exp\left[-\frac{4}{3}\,F^{-1}\left(k^2-E\right)^{3/2}\right]\right\}\ .
\end{equation}
Inserting this into Eq.\ (\ref{renorm}) we obtain, in $D=2$,
\begin{equation}
\label{g2}
g_2(\lambda_R,\mu,E)\sim\frac{1}{\lambda_R}-\frac{1}{4\pi}\,
\ln\left(-\frac{\mu^2}{E}\right)-\frac{i}{4\pi}\,I_2(E)\ ,
\end{equation}
where
\begin{equation}
I_2(E):=\int_0^{\infty}\frac{dk}{\sqrt{k^2-E}}\,
\exp\left[-\frac{4}{3}\,F^{-1}\left(k^2-E\right)^{3/2}\right]\ .
\end{equation}
Consistently with our assumptions on $E$ and $F$ we can
compute $I_2(E)$ using the saddle-point approximation and
obtain
\begin{eqnarray}
I_2(E)&\sim&\int_0^{\infty}\frac{dk}{\sqrt{-E}}\,
\exp\left\{-\frac{4}{3}\,F^{-1}\left[(-E)^{3/2}
+\frac{3}{2}\,(-E)^{1/2}k^2\right]\right\}
\nonumber \\
&=&\sqrt{\frac{\pi F}{8}}\,(-E)^{-3/4}\,
\exp\left[-\frac{4}{3}\,F^{-1}(-E)^{3/2}\right] .
\label{I2}
\end{eqnarray}

In the limit $F\to 0$, the integral $I_2(E)$ vanishes and
$g_2(\lambda_R,\mu,E)$ has a single real and negative zero $E_B$,
corresponding to the energy of the bound state in the absence of the 
external field:
\begin{equation}
E_B=-\mu^2\exp\left[-\frac{4\pi}{\lambda_R(\mu)}\right] .
\label{b2}
\end{equation}
We notice that in $D=2$ a bound state exists --- provided, of course,
$F=0$ --- even if the
renormalized strength of the point-like potential is negative,
in which case one could naively expect the potential to be repulsive.

As the bound state energy must be independent
of the arbitrary scale $\mu$,
one can readily obtain the flow equation for the renormalized
{\em running} coupling parameter, that now reads
\begin{equation}
\lambda_R(\mu)=\frac{\lambda_R(\mu_0)}{1+
[\lambda_R(\mu_0)/2\pi]\ln(\mu/\mu_0)}\ ,
\end{equation}
leading again to asymptotic freedom.

Now, let us consider Eq.\ (\ref{g2}) in the case of a weak field $F$.
Using Eq.\ (\ref{b2}), we can rewrite it as
\begin{equation}
g_2(E_B,E)\sim\frac{1}{4\pi}\left[\,\ln\left(\frac{E}{E_B}\right)
-iI_2(E)\,\right]\ .
\end{equation}
An approximate solution to the equation $g_2(E_B,E)=0$ is given
by $E_0=E_B\left[\,1+iI_2(E_B)\,\right]$; in terms of the
dimensionless variable $\varepsilon=EF^{-2/3}$ we obtain,
cf.\ Eq.\ (\ref{I2}),
\begin{equation}
\label{e02D}
\varepsilon_0\sim
\varepsilon_B\left\{1+i\sqrt{\frac{\pi}{8}}\,(-\varepsilon_B)^{-3/4}
\exp\left[\,-\frac{4}{3}\,(-\varepsilon_B)^{3/2}\,\right]\right\}\ ,
\qquad\quad\varepsilon_B\to-\infty\ .
\end{equation}

Moreover, as in the one- and three-dimensional cases, it is possible
to show that an infinite number of resonances arise as 
solutions to the equation $g_2(\lambda_R,\mu,E)=0$, approaching
asymptotically the half-lines $\arg(E)=0$ and $\arg(E)=-2\pi/3$
when $|E|\to\infty$.


\section{Conclusions}
\label{conclusions}

In this paper we have analyzed the ionization of a very simple (though non-trivial) 
model atom submitted to the influence of a uniform static field.
The model we have considered here is that of a one-electron atom in which 
the Coulomb interaction between the
electron and the nucleus is replaced by an attractive short range
(in fact, point-like) interaction. We have analyzed the problem
in $D=1$, 2 and 3 spatial dimensions.
In spite of its simplicity and of the fact that --- due to the external field ---
the Hamiltonian is not bounded from below, the study of the present model is
far from being academic as it allows to grasp the basic features
of the quantum dynamical behaviour of many realistic physical systems.
In particular, its main prediction is a sensible deviation, in the strong
field regime, from the
naively expected exponential decay law of the survival probability
of the bound state after the external field is turned on.
Actually, more or less important deviations from the exponential
decay law do occur even when the field is weak, specially 
in its short-time behaviour, with the presence of oscillatory transient 
effects (which are more pronounced in $D=1$). 
Such deviations are caused by
the presence of a purely continuous spectrum and the
appearance of an infinite number of resonances once the uniform field is
switched on. 
Deviations from the exponential decay law are also expected for very 
large times; this, however, may be an artifact of the model studied here,
since for more realistic potentials one can prove asymptotic
exponential decay \cite{Herbst}. (On the other hand, as noted before, the
survival probability would be so small when such deviations took place
that they would be practically unobservable.) 


An important development of the present investigation, 
which will be presented elsewhere, 
is the generalization of our analysis to the
additional presence of a uniform magnetic field. In this way, it might be
eventually possible to precisely evaluate the lifetimes of the so called
{\em non-conducting} states --- within the Integer Quantum Hall Effect (IQHE)
conventional terminology --- and to explicitly verify the widely popular
picture according to which the presence of impurities, 
described in the simplest way by point-like attractive wells, 
gives rise to the plateaux formation in the IQHE \cite{QHE}.


\acknowledgments

We thank Roman Jackiw for his useful suggestions.
R.\ M.\ Cavalcanti acknowledges the kind hospitality of
Universit\`a di Bologna and the support from CNPq and FAPERJ.
P.\ Giacconi and R.\ Soldati would like to thank the Institute of Physics
of the Universidade Federal do Rio de Janeiro for the warm
hospitality during a stage of this work.

\appendix


\section{}
\label{A}

In this Appendix we shall sketch the proof of the bound (\ref{INEQ})
on $G^+(E;x,x´)$ in $D=1$. Let us first
examine the asymptotic behavior of $G^+(E;x,x')$
for $|E|\to\infty$ in the sector $-\pi<\arg(E)<-2\pi/3$.
Since $\arg(\rho)\to\arg(E)$ as $|E|\to\infty$, cfr.\ Eq.\ (\ref{rho}),
we have
$|\arg(-\rho)|=|\arg(e^{i\pi}\rho)|<\pi/3$
for $|E|$ large enough, so that \cite{Abramowitz}
\begin{equation}
\label{Ai1}
\Ai(-\rho)\sim\frac{1}{2}\,\pi^{-1/2}(-\rho)^{-1/4}\,
\exp\left[\,-\frac{2}{3}\,(-\rho)^{3/2}\,\right]\ ,
\end{equation}
\begin{equation}
\label{Ci1}
\Ci(-\rho)\sim\pi^{-1/2}(-\rho)^{-1/4}\left\{
\exp\left[\,\frac{2}{3}\,(-\rho)^{3/2}\,\right]+\frac{i}{2}\,
\exp\left[\,-\frac{2}{3}\,(-\rho)^{3/2}\,\right]\right\}\ .
\end{equation}
Inserting Eqs.\ (\ref{Ai1}) and (\ref{Ci1}) into Eq.\ (\ref{G0}),
and dropping the second term in curly brackets in Eq.\ (\ref{Ci1})
since it is negligible compared to the first one, we obtain
\begin{equation}
G_0^+(E;x,x') 
\sim\frac{i}{2}\,F^{-1/3}\,(\rho_{-}\rho_{+})^{-1/4}\,
\exp\left[\,\frac{2i}{3}\left(\rho_{-}^{3/2}
-\rho_{+}^{3/2}\right)\,\right]\ .
\end{equation}
In addition, taking Eq.\ (\ref{rho}) into account, we have
\begin{equation}
\label{rho3/2}
\rho^{3/2}\sim F^{-1}E^{3/2}+\frac{3}{2}\,E^{1/2}x\ ,
\qquad\quad |E|\to\infty\ ,
\end{equation}
so that 
\begin{equation}
G_0^+(E;x,x')\sim\frac{i}{2}\,E^{-1/2}\,
\exp\left(-iE^{1/2}|x-x'|\right)\ .
\end{equation}
Inserting this expression into Eq.\ (\ref{G1})
and using the fact that ${\rm Im}(E^{1/2})<0$,
one can easily show that there is
a positive constant $C$ such that
$|G^+(E;x,x')|< C|E|^{-1/2}$ for $|E|$ sufficiently
large and $-\pi<\arg(E)<-2\pi/3$. 
Notice that Eq.\ (\ref{INEQ}) is a trivial consequence
of this inequality.

Let us now examine the asymptotic behaviour of $G^+(E;x,x')$
in the sector $-2\pi/3<\arg(E)<0$.
For this purpose it is convenient to rewrite
Eq.\ (\ref{G1}) as
\begin{equation}
\label{G2}
G^+(E;x,x')=\frac{G_0^+(E;x,x')+\lambda R(E;x,x')}
{1+\lambda G_0^+(E;0,0)}\ ,
\end{equation}
where
\begin{equation}
\label{G3}
R(E;x,x'):= G_0^+(E;0,0)\,G_0^+(E;x,x')
-G_0^+(E;x,0)\,G_0^+(E;0,x')\ .
\end{equation}
Again, since $\arg(\rho)\to\arg(E)$ as
$|E|\to\infty$, we have $|\arg(\rho)|<2\pi/3$
for $|E|$ large enough,
in which case we have \cite{Abramowitz}
\begin{equation}
\label{Ai2}
\Ai(-\rho)\sim\pi^{-1/2}\rho^{-1/4}\,
\sin\left(\frac{2}{3}\,\rho^{3/2}+\frac{\pi}{4}\right)\ ,
\end{equation}
\begin{equation}
\label{Ci2}
\Ci(-\rho)\sim\pi^{-1/2}\rho^{-1/4}\,
\exp\left[i\left(\frac{2}{3}\,\rho^{3/2}+\frac{\pi}{4}\right)\right]\ ,
\end{equation}
so that
\begin{eqnarray}
G_0^+(E;x,x')&\sim&\frac{i}{2}\,F^{-1/3}\,(\rho_{-}\rho_{+})^{-1/4}
\nonumber \\
& &\times\left\{i\exp\left[\frac{2i}{3}\left(\rho_{+}^{3/2}
+\rho_{-}^{3/2}\right)\right]-\exp\left[\frac{2i}{3}
\left(\rho_{+}^{3/2}-\rho_{-}^{3/2}\right)\right]\right\}\ .
\end{eqnarray}
Using (\ref{rho3/2}) and neglecting the second term in curly
brackets we obtain
\begin{equation}
G_0^+(E;x,x')\sim-\frac{1}{2}\,E^{-1/2}\,
\exp\left\{i\left[\frac{4}{3}\,E^{3/2}F^{-1}
+E^{1/2}(x+x')\right]\right\}\ .
\end{equation}

The asymptotic behavior of $R(E;x,x')$ depends upon the
signs of $x$ and $x'$. Let us first consider the case $x,x'\le 0$. 
If we insert Eq.\ (\ref{G0}) into Eq.\ (\ref{G3}),
and use the identity $f(\rho)\,f(\rho')=f(\rho_{-})\,f(\rho_{+})$,
we obtain ($\varepsilon:=EF^{-2/3}$)
\begin{eqnarray}
\label{G_1}
R(E;x,x')&=&\pi^2 F^{-2/3}\,\Ai(-\rho_{-})\,\Ci(-\varepsilon)
\nonumber \\
& &\times\left[\,\Ai(-\varepsilon)\,\Ci(-\rho_{+})
-\Ai(-\rho_{+})\,\Ci(-\varepsilon)\,\right]\ .
\end{eqnarray}
Furthemore, inserting Eqs.\ (\ref{Ai2})
and (\ref{Ci2}) into Eq.\ (\ref{G_1}) and using
Eq.\ (\ref{rho3/2}) we obtain
\begin{equation}
R(E;x,x')\sim-\frac{1}{2E}\,\exp\left[i\left(
\frac{4}{3}\,E^{3/2}F^{-1}+E^{1/2}x_{-}\right)\right]
\sin\left(E^{1/2}x_{+}\right)\ ,\qquad x,x'\le 0\ ,
\end{equation}
where subdominant terms have been dropped.
A similar analysis shows that
\begin{equation}
R(E;x,x')\sim\frac{1}{2E}\,\exp\left[i\left(
\frac{4}{3}\,E^{3/2}F^{-1}+E^{1/2}x_{+}\right)\right]
\sin\left(E^{1/2}x_{-}\right)\ ,\qquad x,x'\ge 0\ ,
\end{equation}
and that $R(E;x,x')\equiv 0$ for 
$x\le 0\le x'$ or $x'\le 0\le x$. Replacing $G_0^+(E;x,x')$ 
and $R(E;x,x')$ in Eq.\ (\ref{G2}) with their asymptotic 
expressions, and using the inequalities
$|\sin z|\le \exp(|{\rm Im}\, z|)$,\quad
$|{\rm Im}\,E^{1/2}|\le |E|^{1/2}$ and $|x\pm x'|\le |x|+|x'|$,
one can easily derive the bound (\ref{INEQ}).


\section{}
\label{B}

In this Appendix we derive the asymptotic expression of
$\varepsilon_0$ in the weak field limit.
Let us first consider the one-dimensional case.
If we assume that $|\varepsilon|\gg 1$ and
$\arg(\varepsilon)\approx -\pi$, then we may use the
asymptotic expressions (\ref{Ai1}) and (\ref{Ci1})
for the Airy functions $\Ai$ and $\Ci$.
Equation (\ref{EQ}) then becomes
\begin{equation}
(-\varepsilon)^{-1/2}\left\{1+\frac{i}{2}\,
\exp\left[\,-\frac{4}{3}(-\varepsilon)^{3/2}\,\right]\right\}
\approx(-\varepsilon_B)^{-1/2}\ .
\end{equation}
This equation can be solved iteratively. As a first approximation,
one may neglect the second term in square brackets, thus obtaining
$\varepsilon_0\approx\varepsilon_B$.
In order to obtain the
imaginary part of $\varepsilon_0$ one must iterate once more:
replacing $\varepsilon$ in the exponential with
$\varepsilon_B$ and solving the resulting equation, one
finds
\begin{equation}
\varepsilon_0\sim\varepsilon_B\left\{1+i\exp\left[\,-\frac{4}{3}\,
(-\varepsilon_B)^{3/2}\,\right]\right\}\ ,
\qquad\quad\varepsilon_B\to-\infty,\,D=1\ .
\end{equation}
One can derive a systematic expansion in powers of $\varepsilon_B$ if one
includes more and more terms in the asymptotic expansion of
the Airy functions. In particular, the real part of the
resulting expansion for $E_0=F^{2/3}\varepsilon_0$ agrees
with the Rayleigh-Schr\"{o}dinger perturbation series for
the bound state energy when the external field is treated as
a perturbation \cite{Geltman2,Elberfeld}. 

Following the same strategy, we can approximate
Eq.\ (\ref{EQ3}) of the three-dimensional case by
\begin{equation}
(-\varepsilon_B)^{1/2}-(-\varepsilon)^{1/2}-\frac{i}{8\varepsilon}\,
\exp\left[\,-\frac{4}{3}\,(-\varepsilon)^{3/2}\,\right]=0\ .
\end{equation}
We can obtain an approximate solution to this equation using the 
iterative method employed above. This way, we finally arrive at
the result displayed in Eq.\ (\ref{e03D}).
The result for the two-dimensional case is worked out in
Subsection \ref{D=2} and is given in Eq.\ (\ref{e02D}).


\section{}
\label{B/C}

In this Appendix we derive the asymptotic behavior of
$\varepsilon_1$ in the weak field limit, $\varepsilon_B\to -\infty$.
As discussed in Subsection \ref{1D3}, in that limit the r.h.s.\
of Eq.\ (\ref{EQ}) vanishes so that, to the lowest order, one has
$\varepsilon_1\approx -a_1$, where $a_1=-2.33810\ldots$ is the
smallest (in absolute value) zero of $\Ai(z)$. In order to obtain a more
refined approximation, valid for a finite though large value of
$|\varepsilon_B|$, we expand the l.h.s.\ of Eq.\ (\ref{EQ}) in
powers of $x=\varepsilon+a_1$ and, assuming that $|x|\ll 1$,
we truncate the series and solve the resulting polynomial equation
in $x$. The first non-trivial correction to the
imaginary part of $\varepsilon_1$ is obtained when one truncates the
series at $O(x^3)$. In so doing, Eq.\ (\ref{EQ}) is then 
approximated by a quadratic equation in $x$,
\begin{equation}
\label{eq2}
ax^2+bx+c=0\ ,
\end{equation}
where $a=\Ai'(a_1)\,{\Ci}'(a_1)$,
$b=-\Ai'(a_1)\,\Bi(a_1)$, and
$c=-(1/2\pi)\,(-\varepsilon_B)^{-1/2}\ll 1$.
Of the two solutions to Eq.\ (\ref{eq2}),
$x_{\pm}=(-b\pm\sqrt{b^2-4ac})/2a$,
the one with the minus sign must be discarded, as it violates the 
condition that $x\to 0$ as $\varepsilon_B\to-\infty$ ($c\to 0$).
Expanding $x_{+}$ in powers of $c$, we obtain
\begin{equation}
\label{x+}
x_+=-\frac{c}{b}-\frac{ac^2}{b^3}+O(c^3)\ .
\end{equation}
Substituting $a$, $b$ and $c$ with their explicit expressions,
we finally obtain
\begin{eqnarray}
\label{eps1}
\varepsilon_1&=&-a_1+\frac{1}{\Ai'(a_1)\,\Bi(a_1)}\,
\frac{(-\varepsilon_B)^{-1/2}}{2\pi}
\nonumber \\
& &+\frac{{\Ci}'(a_1)}
{\Ai'(a_1)^2\,\Bi(a_1)^3}\,\frac{(-\varepsilon_B)^{-1}}{4\pi^2}
+O\left[(-\varepsilon_B)^{-3/2}\right]\ ,
\qquad\quad\varepsilon_B\to-\infty\ .
\end{eqnarray}
An important consequence of this result is that 
${\rm Im}(\varepsilon_1)\sim(-\varepsilon_B)^{-1}$ for
$\varepsilon_B\to-\infty$.

One could be tempted to apply the reasoning above
to any $\varepsilon_n$, $n\in{\mathbb N}$.
However, there is an important caveat: 
the r.h.s.\ of Eq.\ (\ref{x+}) is a good approximation
to $x_+$ only if $|ac/b^2|\ll 1$, or
\begin{equation}
\label{ac/b2}
\left|\frac{a}{b^2}\right|=
\left|\frac{{\Ci}'(a_n)}{\Ai'(a_n)\,\Bi(a_n)^2}\right|
\ll|c|=2\pi(-\varepsilon_B)^{1/2}\ .
\end{equation}
Using the asymptotic expressions of the Airy functions
and of $a_n$ --- the $n$-th zero of $\Ai(z)$ \cite{Abramowitz} ---
one can show that $|a/b^2|\sim\pi(-a_n)^{1/2}$. Hence,
Eq.\ (\ref{eps1}) is also valid for $\varepsilon_n$, $n>1$
(with the obvious substitution $a_1\to a_n$),
provided $|a_n|\ll|\varepsilon_B|$.
(See Appendix \ref{C} for the asymptotic behavior of $\varepsilon_n$
when $|a_n|\gg|\varepsilon_B|$.)


\section{}
\label{C}

In this Appendix we derive the asymptotic behavior
of the resonances $\varepsilon_n$, $n\ne 0$, which are located very far
from the origin in the complex $\varepsilon$-plane.
Let us first discuss the one-dimensional case.
Assuming that $|\varepsilon|\gg 1$ and
$\theta:=\arg(\varepsilon)\approx 0$, we
are allowed to use the asymptotic expressions (\ref{Ai2})
and (\ref{Ci2})
to the aim of approximating Eq.\ (\ref{EQ}) by
\begin{equation}
\varepsilon^{-1/2}\left[\,\exp\left(
\frac{4}{3}\,i\varepsilon^{3/2}\right)+i\,\right]
\approx(-\varepsilon_B)^{-1/2} .
\end{equation}
If we further assume that $|\varepsilon|\gg|\varepsilon_B|$, we may
neglect the second term in square brackets; the resulting complex
equation
is then equivalent to the following pair of
real equations:
\begin{equation}
\frac{4}{3}\,|\varepsilon|^{3/2}\,\sin\frac{3\theta}{2}
\approx-\frac{1}{2}\,\ln\left|\frac{\varepsilon}
{\varepsilon_B}\right|,\qquad\frac{4}{3}\,|\varepsilon|^{3/2}\,
\cos\frac{3\theta}{2}\approx\frac{\theta}{2}+2n\pi\ ,
\quad\qquad n\in{\mathbb N}\ .
\end{equation}
Assuming $n\gg 1$ and $|\varepsilon|$ large, one can easily
solve these equations, obtaining
\begin{equation}
\theta\sim-\frac{1}{4}\,|\varepsilon|^{-3/2}\,
\ln\left|\frac{\varepsilon}{\varepsilon_B}\right|\ ,
\qquad\quad |\varepsilon|\sim s_n:=
\left(\frac{3n\pi}{2}\right)^{2/3}\ .
\end{equation}
Since $|\theta|\ll 1$, we can write
$\varepsilon=|\varepsilon|\,e^{i\theta}\approx|\varepsilon|\,
(1+i\theta)$, so that
\begin{equation}
\varepsilon_n\sim s_n-\frac{i}{4}\,s_n^{-1/2}\,
\ln\left|\frac{s_n}{\varepsilon_B}\right|\ ,\qquad\quad
n\gg 1,\,D=1\ .
\end{equation}
Next we consider the case $|\varepsilon|\gg 1$ and
$\theta:=\arg(\varepsilon)\approx -2\pi/3$. Using the
very same approximations we readily come to the 
following estimate 
\begin{equation}
\varepsilon_{-n}\sim e^{-2i\pi/3}\left\{s_n+\frac{i}{4}\,s_n^{-1/2}\,
\ln\left|\frac{s_n}{\varepsilon_B}\right|\right\}\ ,\quad\qquad
n\gg 1,\,D=1\ .
\end{equation}
A straightforward generalization of the above treatments to the
basic resonance equation (\ref{EQ3}) in the three-dimensional case
eventually leads to the following asymptotic expressions:
\begin{equation}
\label{asymp3d}
\varepsilon_n\sim s_n-\frac{i}{2}\,s_n^{-1/2}\,
\ln\left(4s_n^{3/2}\right)\ ,\qquad
\varepsilon_{-n}\sim e^{-2i\pi/3}\,\varepsilon_n^*\ ;
\qquad n\gg 1,\,D=3\ .
\end{equation}



\end{document}